\documentclass[useAMS,usenatbib]{mn2e}
\usepackage{graphicx}
\usepackage{subfigure}
\usepackage{amsmath}
\usepackage{epsfig}
\usepackage{epstopdf}
\usepackage{amssymb}
\usepackage[english]{babel}
\usepackage[utf8]{inputenc}
\usepackage{textcomp}

\def\kms{$\textrm{km~s$^{-1}$}$}

\def\kpc{kpc}

\def\H2{H$_{2}$}
\def\HI{H{\sc i}}
\def\HII{H{\sc ii}}
\def\roH2{$\rho_{\textrm{H}_2}$}
     
\def\MH2{M$_{\textrm{H}_2}$}

\def\Ha{H$\alpha$ \,} 
\def\Hb{H$\beta$ \,} 

\def\ii{\,{\sc ii}} 
\def\iii{\,{\sc iii}} 

\def\logoh{$12+\log(O/H)$ \,}

\def\No{\textnumero}

\title[Long-slit spectroscopy of the merging system Arp 270] { Outer regions of the merging system Arp 270\thanks{Based on observations collected with the 6 m
telescope of the Special Astrophysical Observatory of the Russian Academy of
Sciences, which is operated under the financial support of the Science
Department of Russia (registration number 01-43).}}

\author[A. Zasov, A. Saburova, I. Katkov, O. Egorov, V. Afanasiev]
{A. Zasov$^{1,2}$\thanks{Corresponding author\,: {\tt zasov@sai.msu.ru}},
A. Saburova$^{1}$, 
I. Katkov$^{1}$,
O. Egorov$^{1}$, 
V. Afanasiev$^{3}$ \\
$^1$ Sternberg Astronomical Institute, Moscow M.V. Lomonosov State University, Universitetskij pr., 13,  Moscow, 119992, Russia \\
$^2$ Faculty of Physics, Moscow M.V. Lomonosov State University, Leninskie gory 1,  Moscow, 119991, Russia \\
$^3$ Special Astrophysical Observatory, Russian Academy of Sciences, Nizhniy Arkhyz, Karachai-Cherkessian Republic 357147, Russia \\
}

\begin{document}

\pagerange{\pageref{firstpage}--\pageref{lastpage}} \pubyear{2002}

\maketitle

\label{firstpage}

\begin{abstract}

Arp 270 (NGC~3395 and NGC~3396) is the system of two actively star-forming
late-type galaxies in contact, which already have experienced  at least one
close encounter in the past. We performed long-slit observations of peripheric
regions of this  merging system with the 6-m telescope of SAO RAS. Line-of-sight 
velocity distribution along the slits was obtained for gas and stellar
population.  We found that the stellar component of NGC~3395 differs by its velocity from the emission gas component in the extended region in the periphery, which evidences a spatial separation of stars and  gas in the tidally disturbed galaxy. Gas abundances obtained by different methods
demonstrate that both galaxies are mildly underabundant (\logoh $\approx
8.4$) without  significant variations of metallicity along the slits. By
comparing stellar and gaseous masses of galaxies we came to conclusion that
the chemical evolution of gas is badly described by the closed box model. It
allows us to admit that the significant part of interstellar gas was swept out of
galaxies during the preceding encounter(s). A special attention was paid to
the extended kpc-size island of star formation between the galaxies. We have
not found neither noticeable kinematic decoupling of this region from the
adjacent areas, nor any peculiarities of its emission spectra, which evidences
that it was formed recently from the gas of NGC~3395 in the transition region
between the colliding galaxies.

\end{abstract}


\section{Introduction}  

Arp270 = NGC~3395$/$96 = VV246 = KPG 249 -- is tightly interacting pair formed
by two galaxies of comparable luminosities: a spiral galaxy NGC~3395  and
irregular galaxy (Irr or Sm) NGC 3396, highly inclined to the line of sight.
The relative motion of galaxies seems to occur in a plane perpendicular to the
line of sight, because their line-of-sight central velocities nearly coincide
(1625 \kms according to NED\footnote{http://ned.ipac.caltech.edu/}). The
distance between the centres of galaxies is about 10 kpc. Both galaxies
contain neutral and ionized gas and young stars. UV images of galaxies
obtained by the Hubble telescope, revealed in both galaxies a large number of
compact and small (less than 20 pc) regions containing newly formed stars, 
many of which seem to form a gravitationally bound cluster \citep{Hancock}.

There are two features which attract a special attention to this system. First,
here we have the rare case when it is safe to say that the galaxies have
already experienced a strong tidal interaction at least once in the past. Indeed,  \HI\ line
observations reveal the long tidal tails of gas, containing about $4\times10^8$
solar masses of \HI, which extend over a distance of 60 kpc (see \citealt{Clemens}). 
\HI\ data allowed us to obtain a large-scale picture of the
distribution of radial velocities in the interacting system. Numerical
modelling of interactions conducted by \citet{Clemens}, led to
conclusion that the previous convergence of galaxies took place about $5\times10^8$ 
years ago, whereas some tens of millions of years ago the galaxy
once again passed  the position of closest approach. The current splash of
star formation is evidently  associated with this latest convergence. It is worth noting that, as the simulations predict, a strong tidal stripping removes most of the original dark matter haloes of interacting galaxies (\citealt{Libeskind2011}), hence one can expect that the subsequent rapprochement may be more violent due to shallower potential well.  

The second peculiarity of this system is the presence of numerous isolated
islands of star formation, well visible in optical and especially in UV images
in the outskirts of galaxies and between them. In particular, many small blue
stellar regions  are noticeable at the periphery of NGC~3395, in the opposite
side from the companion galaxy, where, apparently, a new tidal tail begins to
grow. Their detailed study is important for elucidating the mechanisms that
promote a formation of stars in the regions of low average density of gas as
well as the birth of dwarf tidal galaxies. The most notable detail is the
extended stellar complex (named as ESC below) of about 10 arcsec $\sim 1$ \kpc\
size between the galaxies, especially bright in the UV image obtained by GALEX
\citep{Smith} (see also Fig.~\ref{arp270_image} in this paper). If it is
gravitationally bound, it can be regarded as the tidal dwarf observed in a
process of formation. In any case, the fact of simultaneous birth of stars in
the extended region situated in a rather rarefied medium between the
interacting galaxies looks curious.

Arp 270 has a fairly complex pattern of internal motions of gas both inside
and outside of galaxies. First optical kinematic measurements made  by
\citet{DOdorico} showed a sign of rotation of the emission gas in both
galaxies. Later, the two-dimensional velocity field in the line $H\alpha$,
was obtained in the framework of the project GHASP \citep{Garrido,Ghasp},
which confirmed the rotation of both galaxies, as well as the presence of
significant non-circular velocity in the system reaching a few tens \kms.
\citet{Zaragoza-Cardiel}  also constructed two-dimensional map of gas
velocities in this system which demonstrates the existence of radial gas
motion in the central region of NGC~3396. Based on the measurements of the
velocity dispersion of emission gas the authors concluded that in both
galaxies star formation is associated mostly with high luminous \HII\ regions,
which apparently are connected with  gravitationally bound clouds.

Masses of \HI\ for both galaxies as well as total mass of \HI\ (including the
gaseous tail), taken from \citet{Clemens}, are given in Table 1 parallel with
the luminosity in $K$-band, corresponding to $K_{tot}$ magnitudes (NED
Data base) for the adopted distance 21.6 Mpc and with the effective oxygen
yields which will be discussed below.

\begin{table}
\caption{Masses of \HI, luminosities and the effective yields for Arp 270}
\label{t1}
\begin{center}
\begin{tabular}{ccccc}
\hline
\\
Object& log $L_K$&$M_H$ & $Y_{eff}\times10^3$  &$Y_{eff}\times10^3$ \\
&($L_\odot$)&($10^9M_\odot$)&M/$L_K$=1&M/$L_K$=0.5\\
\hline
NGC~3395&10.0&1.5&2.4&1.6\\
NGC~3396&10.1&0.92&1.6&1.2\\
total&10.4&2.9&1.9&1.4\\
\hline
\end{tabular}
\end{center}
\end{table}

Our spectroscopic observations were aimed primarily to investigate the
peripheral regions of galaxies Arp~270, their velocity distribution and the
chemical abundance  of gas, as well as to estimate the mean (luminosity-weighted) age of stellar population in order to clarify the evolutionary
status of this system.

\section{Observations and data analysis}

\subsection{Observations and reduction}

\begin{figure}
\centerline{
\includegraphics[width=0.5\textwidth]{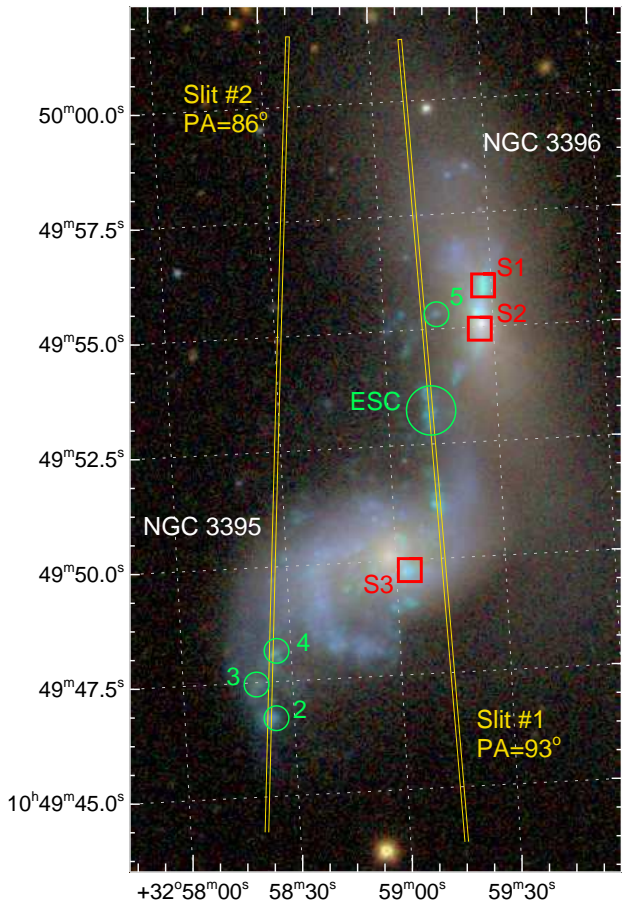}}
\caption{Composite colour map of Arp~270 based on $gri$-band images from SDSS. 
Long slit orientations are shown in yellow. Green circles mark star-forming
regions which are discussed in the text. Red squares correspond to the positions 
of the SDSS aperture spectra.}
\label{arp270_image}
\end{figure}

Long-slit observations of galaxy system Arp~270 were performed in 2013 February
  with a new universal spectrograph SCORPIO-2 \citep{scorpio2} mounted at
the prime focus of the 6-m Russian telescope BTA at Special Astrophysical
Observatory of the Russian Academy of Sciences.  Arp~270 was observed with
1 arcsec-width long slit for two different orientations that are shown in
Fig.~\ref{arp270_image}. Hereafter, the slit with the positional angle
PA=93$^\circ$ is labelled as Slit~\No 1 and the second slit with PA=86$^\circ$
is labelled as Slit~\No2 . The spectra were obtained with volume-phase
holographic grism VPHG1200@540 which provides spectral coverage from 3600
to 7200\AA. We used CCD chip E2V CCD42-90  $2048\times4600$ in the $1\times2$
binning mode which provided a spatial scale of 0.357 arcsec pixel$^{-1}$ and
spectral sampling $\approx1.6$\AA\ pixel$^{-1}$. Exposition time and averaged
atmospheric conditions are shown in the observational log in
Table~\ref{tbl_obs}.

\begin{table}
\caption{Observational log}
\label{tbl_obs}
\begin{center}
\begin{tabular}{cccc}
\hline
\\
Date & $PA_{slit}$ & $T_{exp}$  & Seeing \\
     & (deg)       & (sec)      & ($arcsec$)\\
\hline
09.02.2013&93&3900&4\\
09.02.2013&86&3600&2.5\\
\hline
\end{tabular}
\end{center}
\end{table}

The primary data reduction steps comprised of bias subtraction, 
flat-fielding and cosmic ray hit removal by using Laplacian filtering technique
\citep{lacosmic}. We constructed the wavelength solution using  the He-Ne-Ar lamp
by fitting the position of lines by fifth- and fourth- order two-dimensional
polynomial across and along the dispersion correspondingly, followed by
linearization of spectra. The uncertainties of the wavelength solution
$<0.15$\AA\ with the mean error $0.08$\AA.

To estimate a night sky spectrum, we used the peripheral regions of the
slits where the contribution of galaxies is negligible. Then, the peripheral night
sky spectrum  was  transformed into the Fourier space and extrapolated on to
the galaxy position by using polynomial representation at a given frequency
and after that the inverse Fourier transformation was performed. This night
sky subtraction technique allows us to take into account a smoothed variation of
the line spread function (LSF) of the spectrograph. Unfortunately we could not
applied the previously developed advanced sky subtraction technique
\citep{skysubtraction} because it requires the reference twilight sunsky
spectrum which was not obtained in the same observational nights. 
Final reduction steps were the summing
of separately reduced object exposures and flux calibration using the spectrum
of spectrophotometric stellar standard HZ4.

For the accurate extraction of stellar and emission-line kinematics
(see below), we included into the model the information about LSF. Despite the
absence of twilight spectrum, we mapped the LSF by fitting the Gauss-Hermite
functions up to fourth order against the lines of the arc lamp spectrum used for
the  wavelength calibration.

\subsection{Stellar population properties}

We failed to extract the continuous radial profiles of stellar
population properties along the slits due to the low signal-to-noise ratio ($S/N$).
Instead, we summarized long-slit spectra in two bins for two sections of  Slit
\No1 and in one bin for Slit \No2. The region of summation of spectra and
the corresponding SNR are shown in Table~\ref{tbl_stellar_populations}. 

Arp~270 is an actively star-forming system, therefore the contribution of young
stellar population to the galaxy spectra is essential. We
utilized the  high-resolution ($\approx0.3$\AA) stellar population models
produced by Starburst99 syntheses code \citep{Leitherer1999_sb99,
Vazquez2005_sb99}. Recently, Starburst99 code was optimized for inclusion of
the  intermediate-age and old stellar populations by incorporation of the
Padova evolutionary stellar tracks into the code that allowed us to extend the
modelling to ages older than $\approx 1$ Gyr and stellar  masses below
$0.8M\odot$ \citep{Vazquez2005_sb99}. We used the model templates computed
for the Padova tracks, the stellar synthetic library presented by
\citet{Martins2005},  the instantaneous star formation history (SSP - single
stellar population) and the Kroupa initial mass function \citep{Kroupa2002}.
The final model grids are parametrized by age $T$ and metallicity [Z/H].

Stellar population parameters were derived by using a full-spectral fitting
technique \textsc{nbursts} \citep{nbursts_a,nbursts_b} in the special $\chi^2$
scanning mode. This mode consists of the simultaneous non-linear,
least-square fitting of parameters specified the internal kinematics in the
Gauss-Hermite form \citep{gausshermite}, parallel with the stellar population
template pre-convolved with instrumental LSF and
multiplicative continuum. The multiplicative continuum was represented by
cubic spline function which connects 10 node points uniformly distributed
along the wavelength. Note that the continuum fitting is of special importance
because it takes into account a possible internal dust reddening as well as
residual spectral slope variations due to errors in the assumed instrumental
response.

\begin{figure*}
\centerline{
\includegraphics[width=0.33\textwidth]{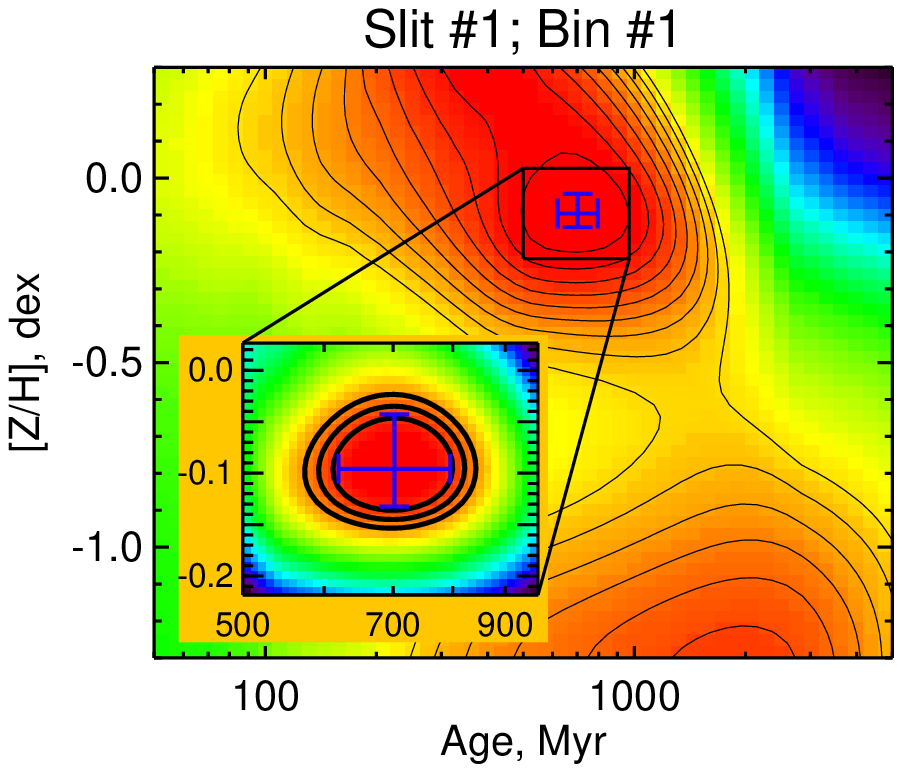}
\includegraphics[width=0.33\textwidth]{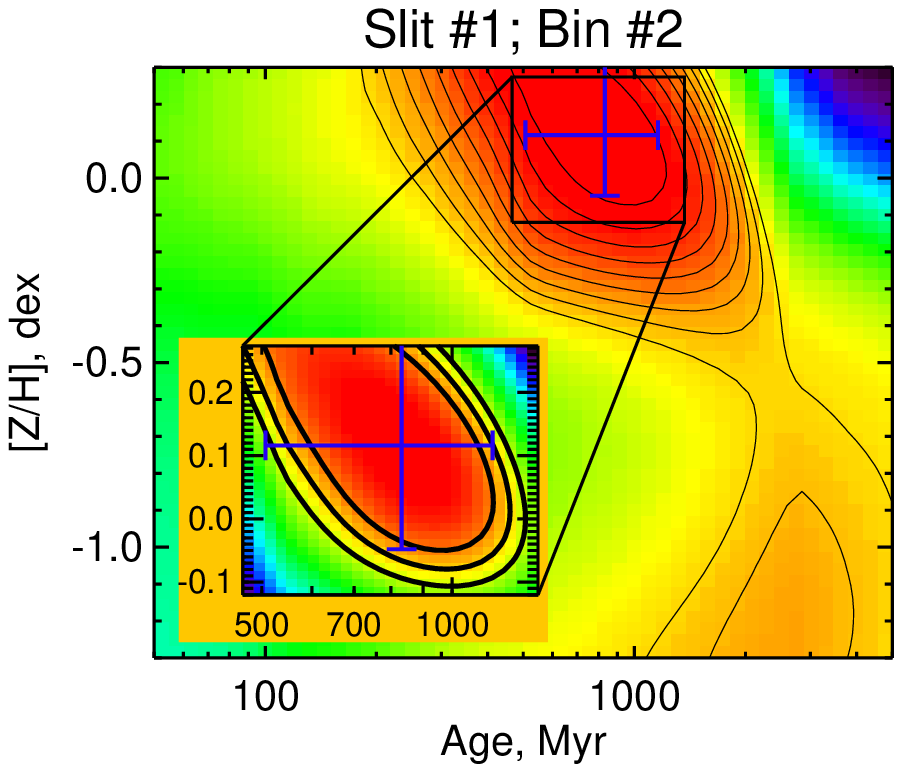}
\includegraphics[width=0.33\textwidth]{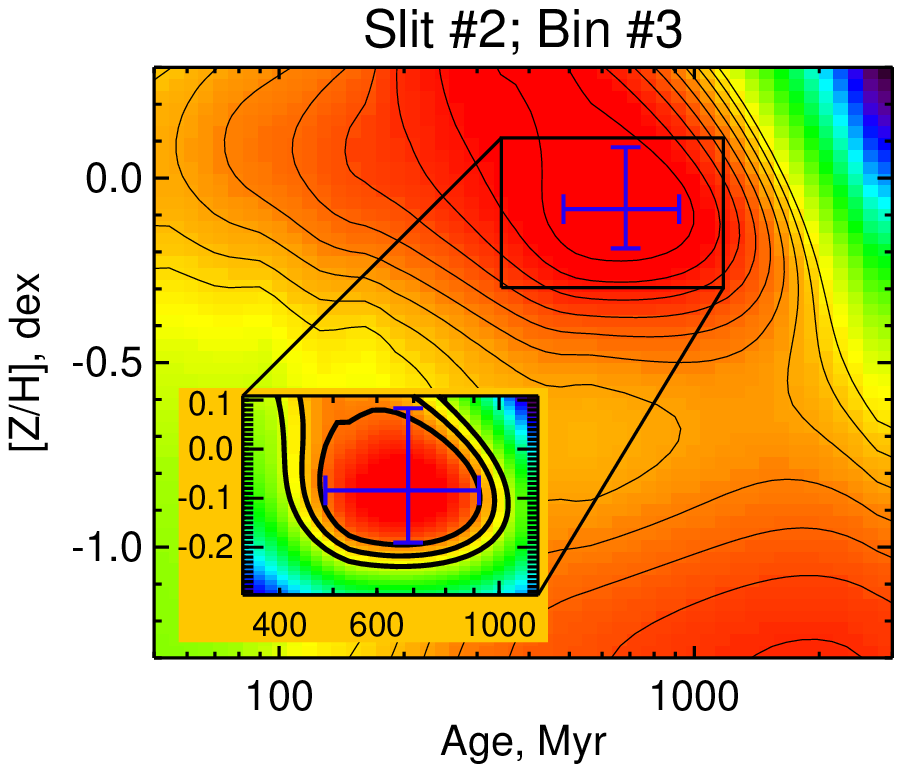}	
}
\caption{$\chi^2$-maps in the age-metallicity plane for different binned
regions along slits. Black contours at the zoomed subregion of $\chi^2$ map
correspond to $1\sigma$, $2\sigma$, $3\sigma$ confidence levels. Blue error
bars in both maps are computed by using $1\sigma$ level contour.}
\label{fig_chi2maps}
\end{figure*}

In this work, we proceeded the fitting of radial velocity and multiplicative
continuum only for a set of fixed values of ages, metallicities and velocity
dispersions. For each combination of these model parameters
($T$,[Z/H],$\sigma$), we computed the optimal values of other parameters parallel with
$\chi^2$ value that finally produced $\chi^2$-cube in the parameter space
$T-[Z/H]-\sigma$. For any acceptable values of $T$ and $[Z/H]$ the minimal $\chi^2$ was
achieved at the lowest velocity dispersion of about 20 \kms\ for every bin.
Taking into account the instrumental resolution ($\sigma_{inst}=115$ \kms), we
can suggest that the stellar velocity dispersion for every bin does not exceed
0.5$\sigma_{inst}\approx 60$ \kms. The $\chi^2$ maps for each bin are shown at
Fig.~\ref{fig_chi2maps}. It shows that for each bin there exists a global minimum
in the $\chi^2$ maps. There is also a substantial age-metallicity degeneracy
for Bin~\No2 (see zoomed subimages at Fig.~\ref{fig_chi2maps}) that appears as
the noticeably stretched and inclined isolines probably caused by the low signal-to-noise 
ratio ($S/N=5.5$ per pixel as about 5100\AA).  The best-fitting values of
stellar population parameters are presented in
Table~\ref{tbl_stellar_populations}, the parameter errors there correspond to the
range covered $1\sigma$  level at the $\chi^2$ maps (see
Fig.~\ref{fig_chi2maps}).

Note that the retrieved stellar population age $T$ may not correspond to
the epoch of formation of the entire bulk of stars because the integrated spectra
for every bins keep fossil record of young, intermediate-age  and old stars.
The obtained SSP-equivalent  stellar population parameters are rather
luminosity-weighted ones for stars of all ages and metallicities.
It means that the  SSP-equivalent estimates
are biased to properties of younger stellar populations.

\begin{table*}
\caption{Stellar population properties}\label{tbl_stellar_populations}
\begin{center}
\begin{tabular}{cccccc}
\hline\hline

Slit & Bin & Position & $S/N$ & $T_{SSP}$   & [Z/H]$_{SSP}$ \\
     &     & (arcsec) &       & ($10^6$ yr) & (dex)\\
\hline
\No~1 (PA=93$^\circ$) & \No~1 & {[{-52},-16]} & 20.3 & $702^{+93}_{ -82}$  & $-0.10^{  +0.05}_{-0.04}$\\[1ex]
\No~1 (PA=93$^\circ$) & \No~2 & {[{ 16}, 77]} & 5.5 & $832^{+326}_{-325}$  & $ 0.12^{ +0.19}_{-0.16}$\\[1ex]
\No~2 (PA=86$^\circ$) & \No~3 & {[{-22}, 31]} & 7.0 & $683^{+235}_{-199}$  & $-0.08^{ +0.17}_{-0.11}$\\
\hline\hline
\end{tabular}
\end{center}
\end{table*}

\subsection{Line-of-sight velocities of stars and gas}

\begin{figure*}
\centerline{
\includegraphics[width=0.5\textwidth]{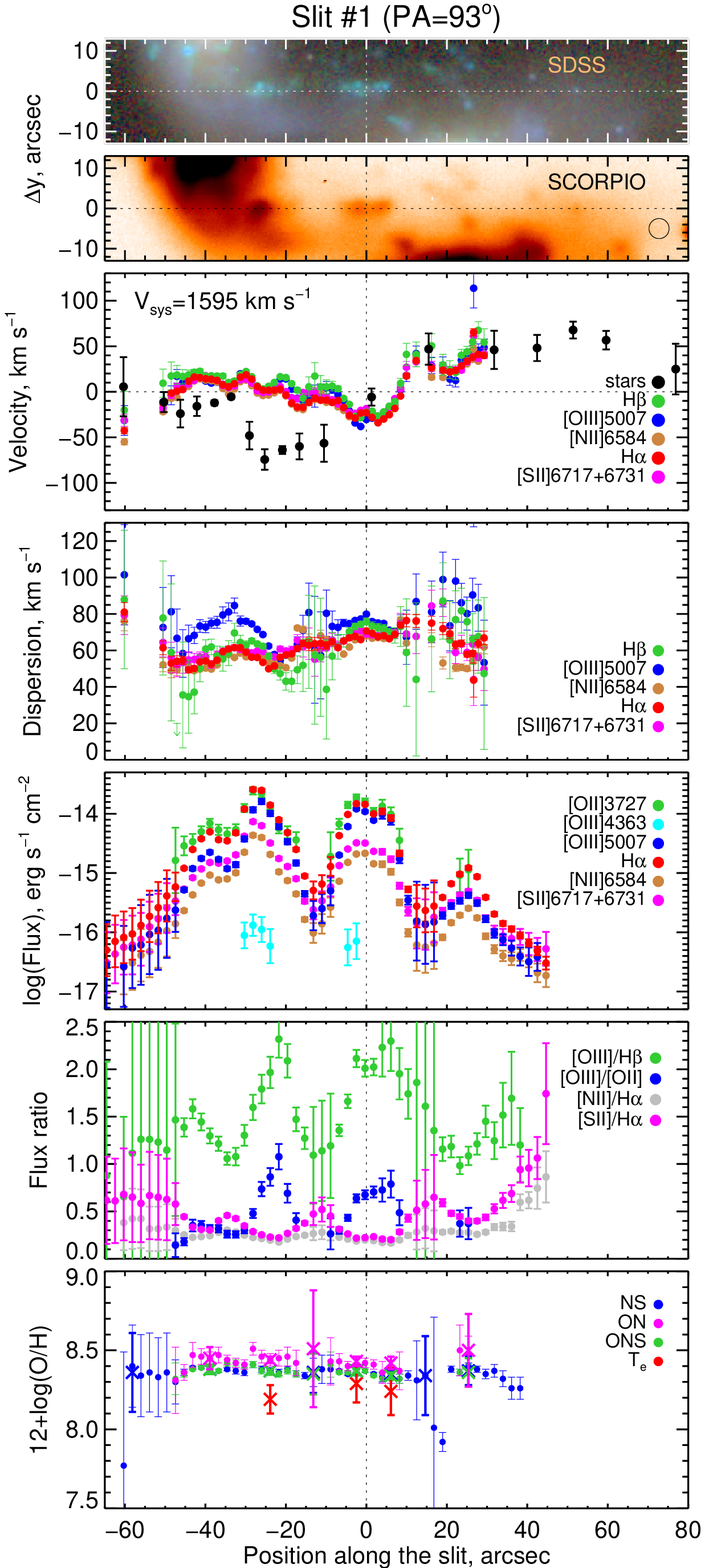}
\includegraphics[width=0.5\textwidth]{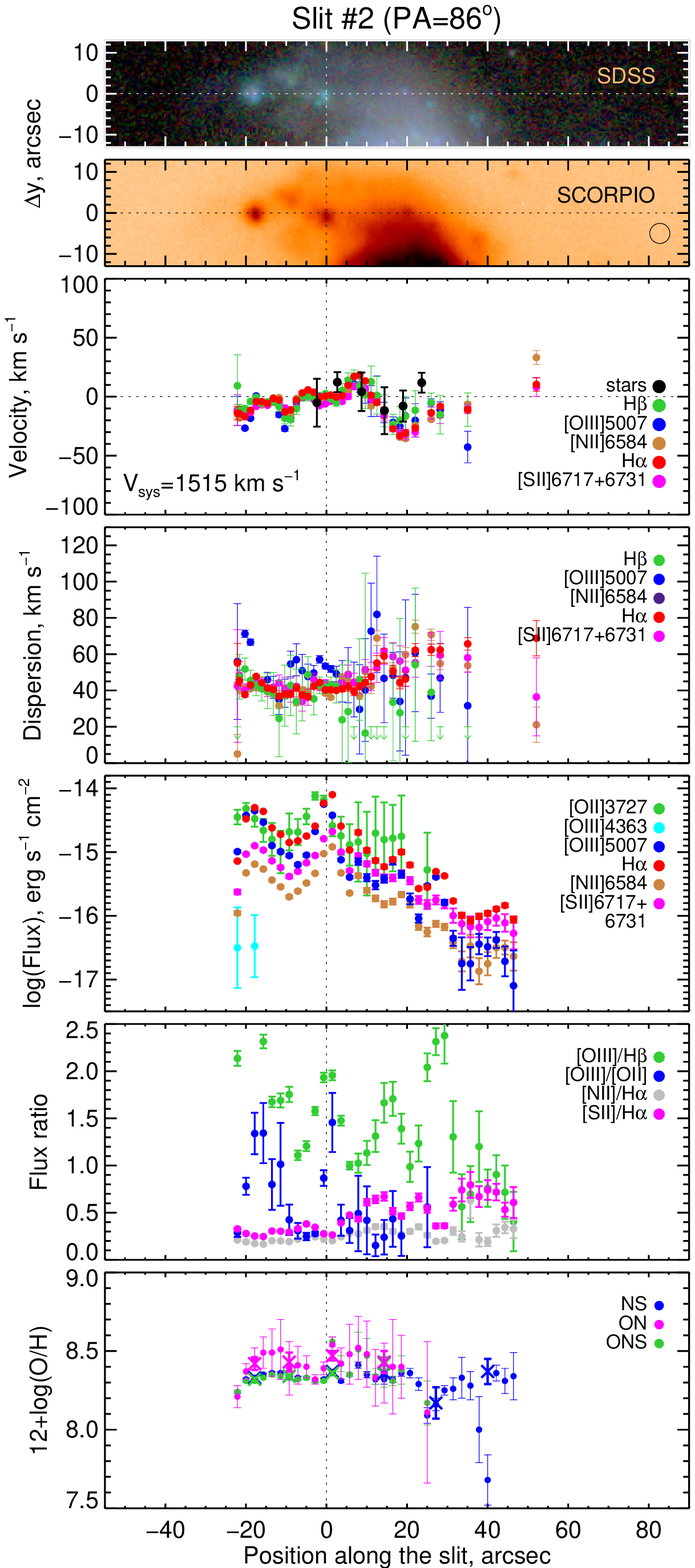}}

\caption{Radial profiles of the line-of-sight velocity and velocity dispersion and distribution of emission-line fluxes, flux ratios and oxygen abundances 
measured for both slits; (\textit{left}  Slit~\No1 (PA=93$^\circ$),
\textit{right} - Slit~\No2 (PA=86$^\circ$). Two top
panels present the reference images taken from SDSS and obtained at SCOPRIO in
the image mode. Black circle describes the atmospheric seeing. Next two panels
reproduce the line-of sight velocities and velocity dispersions. Zero level
of velocities  ($V_{sys}$) are given in the panels. Black symbols correspond
to stellar kinematics, colour circles to different emission lines. The zero-
points on the $x$-axes correspond to the most bright regions at the
slits. Bottom three panels present the fluxes of  measured
emission lines, line flux ratios and oxygen abundances
calculated by different methods (see details in
Section~\ref{section_gas_abund}). Crosses in the bottom panels indicate the results obtained using the fluxes integrated over the regions with larger binning.}

\label{profiles_kin}
\end{figure*}

To extract the profiles of the line-of-sight stellar velocities along the
slits, we divided spectra on smaller spatial bins used previously for
determination of stellar population parameters  by applying the adaptive
binning of the long-slit spectra to achieve minimal required value S/N=3.
Then, we fitted the binned spectra by high-resolution Starburst99 models with
the fixed parameters of velocity dispersion and stellar population by varying only
the line-of-sight velocity and multiplicative parameters. Stellar
velocity dispersions were taken as 20 \kms, which agrees with global minima
in the  $\chi^2$-cubes. The ages and metallicities were fixed as constant
values, found from the analysis of spectra in large bins, described
above. The uncertainties of the velocities were estimated by Monte Carlo
simulations for each spatial bin. The radial velocity profiles are shown by
black symbols in Fig.~\ref{profiles_kin}.

The emission-line spectral component was obtained by subtracting the 
best-fitting stellar population model from the observed spectrum. This step
provided a pure emission spectrum uncontaminated by stellar absorption lines
that is important for the Balmer lines. Then we fitted emission lines with
Gaussian profiles  pre-convolved with instrumental LSF in order to determine the 
line-of-sight velocities of the ionized gas and emission-line fluxes. The velocity
profiles of emission lines are shown for both slits in
Fig.~\ref{profiles_kin}. The analysis of emission-line ratios and ionized gas
abundances is presented below in Section~\ref{section_gas_abund}.

\subsection{Chemical abundance of ionized gas}\label{section_gas_abund}

Emission-line flux measurements allow us to get the information about the
gas ionization state and  its chemical composition. The distributions of
emission-line fluxes and their ratios along the
slits are shown in Fig.~\ref{profiles_kin}. The corrections for reddening were
calculated from the Balmer decrement for all line fluxes. As it follows from
Fig.~\ref{profiles_kin}, the highest values of line ratios [O\iii]/[O\ii] as
well as [O\iii]/H$\beta$ in the spectrum for  PA=93$^\circ$ (Slit \No1)
take place in the region of star-forming `island' (ESC) and slightly
eastward from the compact bright star formation site at the position -25
arcsec along the slit. It gives evidence that there we have the highest
ionization degree, as well as the most hard emission in comparison with
the adjacent regions. In general, the ionization state smoothly
decreases along Slit \No1 from the east to the west edge.


The spectrum with PA=86$^\circ$ (Slit \No2) reveals the enhanced ratio of
[O\iii]/\Hb\ in the most bright detached \HII\ regions. Their ratios of
[N\ii]/\Ha\ and [S\ii]/\Ha\ (0.2-0.3) are almost equal to
those for the main galaxy body and correspond to the photoionization
mechanism of the emission-line excitation. The enhanced line ratios
[N\ii]/\Ha\  and [S\ii]/\Ha\ are mostly observed in the faintest
and evidently most rarefied regions.

\begin{figure}
\includegraphics[width=\linewidth]{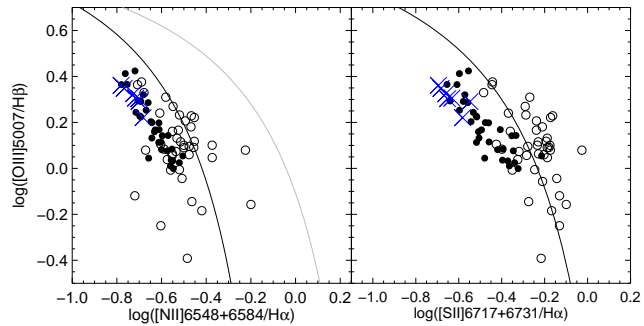}
\caption{Diagnostic diagrams for studied bins along both slits. Open circles
denote faint bins with $\mathrm{F(H\alpha) < 10^{-15} erg s^{-1} cm^{-2}}$,
while filled circles correspond to bright regions  $\mathrm{F(H\alpha) >
10^{-15} erg s^{-1} cm^{-2}}$. Position of ESC is marked by blue crosses. Black
curved line separates the pure photoionization regions from the regions with the
shock excitation (following \citealt{Kewley2006}). Regions with the mixed ionization
mechanism lay between the black and grey lines in the left diagram.}
\label{fig:bpt} 
\end{figure}

In Fig.~\ref{fig:bpt} the diagnostic diagrams of the line fluxes ratio
[O\iii]/\Hb\ over [N\ii]/\Ha\ (left-hand panel) and over
[S\ii]/\Ha\ (right-hand panel) are shown. The curved lines in the diagrams 
separate the regions of shock ionization (above) and photoionization 
(below; see \citealt{Kewley2006}).  In the left-hand panel of Fig.~\ref{fig:bpt} several `composite' 
regions with the mixed excitation  mechanism are located between the grey 
and black lines.

In the diagnostic diagrams we marked the regions with different intensity
levels  by different symbols: bright regions ($\mathrm{F(H\alpha) > 10^{-15}
erg\ s^{-1} cm^{-2}}$)  we denoted with filled circles, while faint regions
($\mathrm{F(H\alpha) < 10^{-15} erg\ s^{-1} cm^{-2}}$)  are marked by the open
circles. Crosses in the diagnostic diagrams mark the positions of the ESC region.
One can see that for all studied bright regions  (including ESC) a
photoionization is the dominant mechanism of excitation,  while the presence
of the shock waves appears in some faint diffuse regions only.

For the detailed investigation of gas chemical abundance (metallicity), we have
to know the electron temperature of the emission  regions. Electron
temperatures $T_e$ of high and  low excitation regions may be found from the
measurements of emission lines  [O\iii] 4363\AA\ and [N\ii] 5755\AA\ fluxes
sensitive to $T_e$ (see e.g. \citealt{Pilyugin2009}). However,  these lines are
faint, and we were able to get the accurate measurements of these lines only
for several regions. The location of these regions along the slit and
the corresponding oxygen abundance are shown in Fig.~\ref{profiles_kin}
and denoted with red crosses.


Applying the equations 
from \cite{Izotov2006}, we estimated the abundances of oxygen, nitrogen, sulphur, argon, neon, 
iron and chlorine using the direct $T_e$-method for regions with available $T_e$ estimations. 
The median obtained values are: 
$\mathrm{12+\log(O/H) = 8.22\pm0.13}$, $\mathrm{\log(N/O) = -1.10\pm0.14}$, 
$\mathrm{\log(S/O) = -1.77\pm0.16}$, $\mathrm{\log(Ne/O) = -0.78\pm0.26}$, 
$\mathrm{\log(Ar/O) = -2.53\pm0.15}$, $\mathrm{\log(Fe/O) = -2.01\pm0.14}$, 
$\mathrm{\log(Cl/O) = -3.55\pm0.15}$. These calculated chemical abundances relative to oxygen 
abundance are in good agreement with the mean values for other spiral galaxies~\citep{Izotov2006}.

As far as the photoionization excitation mechanism dominates in the
investigated regions, we may conclude that most probably emission lines are produced by non-resolved \HII\ regions.  In this case we are able to use the empirical methods for metallicity
estimations calibrated by the data for \HII\ regions with
available accurate measurements of $T_e$. We applied three of these methods to
our data:

\begin{itemize}


\item ONS-method \citep{Pilyugin2010}, that uses the flux ratios of
[O\ii]3727+3729, [O\iii]4959+5007, [N\ii]6548+6584, [S\ii]6717+6731 lines to \Hb

\item ON-method (Pilyugin et al. 2010), that uses the ratios of [O\ii]3727+3729, 
[O\iii]4959+5007 line fluxes to \Hb

\item NS-method \citep{Pilyugin2011}, that does not demand the knowledge of intensity of emission 
line [O\ii]3727+3729. Instead, it uses the line flux ratios of [O\iii]4959+5007, 
[N\ii]6548+6584, [S\ii]6717+6731 to H$\beta$.

\end{itemize}

The distribution of oxygen abundance along the slits, found by different
methods is shown in Fig.~\ref{profiles_kin} (bottom panels). ON, NS and ONS methods give
similar results. The mean metallicity $\mathrm{12+\log(O/H)=8.35\pm0.11}$ is
slightly higher than that obtained by `direct' $T_e$-method for
regions with available $T_e$ estimations.

In addition, we found three spectra for studied galaxies in the SDSS
archive which positions are shown in Fig.~\ref{arp270_image}. Unfortunately,
their spectral range does not cover the [O\ii]3727 line that is necessary for
oxygen abundance estimation by $T_e$-method. We were able to measure the
metallicity from these spectra only using the NS method. The values of
$\mathrm{12+\log(O/H)}$ obtained were 8.37, 8.44 and 8.34 corresponding 
to S1, S2, S3 marks in Fig.~\ref{arp270_image}, that are very close
to the measurements described above for our spectra.

\begin{figure}
\includegraphics[width=\linewidth]{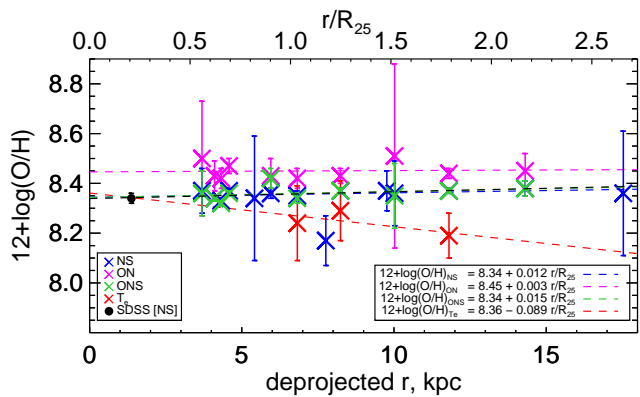}

\caption{Radial distribution of oxygen abundance in the NGC~3395 galaxy obtained by different methods (see text). Crosses denote the same as in Fig.~\ref{profiles_kin}, black point denotes the result obtained with SDSS data. The linear robust fits of oxygen abundance distribution for different methods are shown with dashed lines of corresponding colour.}
\label{fig_abund_rad} 
\end{figure}

The main feature of the obtained oxygen abundance distribution is the  absence
of metallicity gradient. To demonstrate it, we give the plot of radial distribution of metallicity in Fig~\ref{fig_abund_rad}. In this figure the mean values of gas metallicity $12+\log\mathrm{(O/H)}$ are plotted against  their deprojected  distance from the centre of NGC~3395 galaxy (in terms of kpc and of $R_{25} = 6.6$~kpc fraction). We made the deprojection assuming inclination angle $i=57.8^\circ$ and major axis position angle PA=35$^\circ$ (from HyperLeda\footnote{http://leda.univ-lyon1.fr/} data base; \citealt{Makarov2014}).  As it follows from Fig.~\ref{fig_abund_rad},  the gas
metallicity is almost the same for all investigated regions being close to
those found from the SDSS data for the inner galaxy. The metallicity gradients obtained for each methods are very low or negligible. Largest gradient ($-0.089\pm0.091$~dex/$R_{25}$) is given by $T_e$ method using only three available data points with large uncertainties. But even in this case, the gradient is low in comparison with normal galaxies. Of course, our observational data do not provide a dense sample of oxygen abundance measurements at different galactocentric distance and 3D spectroscopy observations of this system are needed to strength this conclusion. Nevertheless, if the metallicity  gradient was high, we would have found it.

\section{Discussion}

\subsection{Stellar and gaseous kinematics}

The distribution of line-of-sight velocity along the Slit \No1
(Fig.~\ref{profiles_kin}, \textit{left}-hand panel) which runs approximately
parallel to the line connecting the centres of galaxies, confirms the opposite
directions of radial velocity gradients of two galaxies found in the 
earlier papers  (see the Introduction). There is a good concordance between the 
velocity measurements for different emission lines. Significant irregularities
of velocity distributions reveal the existence of non-circular motions of gas 
in a kpc-scale; however,  they are not as large as those found in the inner 
parts of galaxies. Most notable changes of the velocity coincide with the bright
emission areas, although the most extended emission region in the Slit \No1
(ESC)  does not manifest itself on the velocity profile. There is a steep
velocity gradient in the region between galaxies (Slit \No1) accompanied by grow of
velocity dispersion -- evidently as the result of direct contact of gaseous
systems of two galaxies. Note that the brightest emission region (ESC) lies
at the beginning of this transition region between galaxies. Similarly, steep gradient is also
presented in the second cut (Slit \No2) where the slit crosses the outer regions
of NGC~3395 near its major axis. It evidences strong ($\approx$50
\kms) perturbations of the velocity field in the region where the short tidal
tail seems to grow.

Stellar line-of-sight velocities agree with the gaseous ones with the
exception of the extended region of about 3 kpc length between the main body
of NGC~3395 and the emission region ESC (Fig.~\ref{profiles_kin}). There the
stellar velocities are up to 50 \kms\ lower than the emission gas velocities
observed along the same lines of sight. There is no such discrepancy along the
Slit \No2. Although the surface brightness of this region is rather low, small error bars demonstrate a reliability of estimates. The low-velocity stellar component hardly may be ascribed to the second galaxy, because its adjacent side is the receding one, having a lager line-of-sight velocity. Gas velocity field of both galaxies reveals a strong non-circular motion, although a rotational pattern of the velocity field of NGC~3395 is evident (\citealt{Zaragoza-Cardiel}). Measured velocities  of emission gas (\citealt{Zaragoza-Cardiel}) and \HI\ (\citealt{Clemens})  in this area  are several tens \kms\  above the velocity of galaxy centre due to the disc  rotation,  whereas our measurements give the  velocities of stars which are several tens \kms\ lower than that of gas. The most natural explanation of the discrepancy is that the gas and stars are spatially separated there because of strong disc tidal distortion, so that the stellar disc is more face-on oriented than the gas layer in the region crossed by the slit, and, as a result, its rotational velocity component is small. Certainly, to get a detailed picture of stellar kinematics, we need a two-dimensional stellar velocity field, which is absent now.

We compared the velocity range estimated in this paper with that of the velocity field of \cite{Zaragoza-Cardiel}. For the Slit~\No 1, we got the velocity variance of $\Delta v \approx 1565-1655$ \kms\ for ionized gas ( $1535-1655$ \kms\ for stars). The velocity range given by \cite{Zaragoza-Cardiel} in the slit region is $\Delta v \approx 1590-1650$ \kms. This dispersion is lower than it follows from our observation; however, the mean velocities of ionized gas agree satisfactorily. For Slit~\No 2 the situation is similar. The estimated velocity range is $\Delta v \approx 1485-1535$ \kms\ in this paper. The velocity field of \cite{Zaragoza-Cardiel} gives $\Delta v \approx 1490-1550$ \kms\ for this region. Note however that the velocities found from \HI\ observations with low spatial resolution (\citealt{Clemens}) are systematically higher than our estimations at  about 60 \kms, although the signs of velocity gradients for both slits are the same.

Fig.~\ref{profiles_kin} also presents the velocity dispersion of gas
calculated from the instrumental profile-corrected line widths. Our
measurements confirm a high velocity dispersion of ionized  gas
obtained from different emission lines ($\approx 50-70$ \kms), previously
found  by \citet{Zaragoza-Cardiel} for the main bodies of both galaxies. Note
that the velocity dispersion in excess of 40 \kms\ almost never occurs in
normal mildly star-forming galaxies, instead it is usually observed only in galaxies with the
highest luminosity in the \Ha\ line. In galaxies with moderate star
formation rate, the observed dispersion usually lays in the range of $20-40$
\kms\ (see e.g. \citealt{Moiseev2014}). In the case of Arp~270, the highest
values of velocity dispersion (up to 80 \kms) accompanied by the high excitation
line O[\iii]5007 are observed in the regions most close to the centres of galaxies, which
evidently  reveals the presence of shock waves there.

It is worth mentioning that both slits avoid the central parts of galaxies,
where  star formation is most active. A cause of high velocity
dispersion of gas may be either internal velocity dispersion inside of large
\HII\ regions, or the spread of relative velocities of separate star-forming
regions on the scale of  linear spectral resolution (hundreds of parsecs), or
both. The fact that the velocity dispersion remains high between the
observed \HII-regions gives evidence against the first version. Most probably,
it reflects a small-scale turbulence due to gas perturbations
feeded by tidal forces. It is worth noting that the velocity dispersion
of the cold gas (\HI) obtained by \citet{Clemens} for the angular
resolution of $\sim 21$ arcsec is about 27 \kms\ (for NGC~3395),  which is also higher than the
typical values 10-15 \kms\ for star-forming galaxies. A large velocity variance
along the slit in the eastern part of NGC~3396, where, apparently, a tidal
tail forms, evidences that the gas velocity field is very complex there due to
tidal streams of gas.

As for the stellar velocity dispersion, we failed to measure it reliably even
using the binnning to combine the adjusted regions. We may conclude however
that it is highly probable that it does not exceed 50 \kms.

\subsection{Oxygen abundance and yield}

The most important feature of the abundance distributions along the slits is
the absence of significant gradients of metallicity, although the better spatial coverage of the spectroscopy is needed to make firmer conclusion. Note that  the low radial
gradient of metallicity is quite common situation for the outer \HII-regions
and for interacting systems \citep{Werk2010, Werk2011, Sanchez2014,
Miralles2014}. Among the proposed mechanisms of smoothing, the abundance
gradient well appropriate for our case, is the radial mixing of metals caused by tidal torques (\citealt*{Rupke2010}). The most puzzling
circumstance in the case of Arp~270 is that, unlike   spiral galaxies, where
the abundance gradient usually flattens in the outer disc, in our case the
abundance remains approximately the same all around the system inspite of the
significant range of radial distances covered by the slits (see Fig. 5). Different methods
of the abundance estimates give the oxygen abundance  \logoh $\approx8.4$. Spectral data presented by  Sloan Survey  after the data processing,
similar to that we used above,  give nearly the same abundance \logoh $\approx
8.4$ for the central parts of both galaxies  (see Section ~\ref{section_gas_abund}) as we observe in the outskirts of the system.

It is essential that we estimated the metal abundance of gas continuously
along the slit, not  for selected high luminous \HII\ regions. Hence, the
metallicity we found is not a result of self-enrichment of massive star-forming
clouds, which  shows that the elements are more or less homogeneously
distributed in the interstellar medium. It is evident that there is no enough
time for gas mixing since the previous encounter of galaxies if to assume that
the metal mixing spreads out with the turbulent velocity of gas. To get the
effective missing, we need  either the extended gas flows or the mixing via a
hot gas phase (see the discussion by \citealt{Werk2011}).

Taking solar  oxygen abundance as  \logoh $= 8.7$ \citep{Asplund2009},  we
find that the oxygen abundance in Arp~270 is about 1/2 solar value, which is
close to the metallicity of  gas in spiral galaxy M33. According to 
\citet{Magrini2010}, the mean abundance O/H in M33 varies between 8.4
(inner part of the disc) and 8.2, that is the gas metallicity $1/2 - 1/3$ of
solar value.  Indeed, the size and luminosity of both galaxies in Arp~270 are
not very different from those for M33. The difference, however, is noticeable
when comparing the relative mass of the gas with respect to the stellar mass
of galaxies.  Total mass of \HI\ in M33 is $3.2\times10^9 M\odot$
\citep{Corbelli2003}, which is about a half of total mass of the disc
according to the mass model of this galaxy \citep{Saburova2012}, that is the
gas and stellar components of the disc are comparable by mass. On the other
hand, for Arp 270 this ratio is much lower. Let us assume that $M/L_K$ of stellar
population lays between 0.5 and 1.0 solar units, as stellar population models
predict for star-forming galaxies (see e.g. \citep{Bell2001, Bell2003}). Then
from the data presented in Table~\ref{t1}, it follows that  the gas-to-star mass
ratios  $M_{gas}/M_*$ for the system as whole is 25-12.5\% (the latter value
is for $M/L_K$=1).  Here, the presence of helium is taken into account. We neglect molecular gas in galaxies: its mass is usually much less than the mass of \HI. Indeed, direct measurements of CO emission for the larger of the galaxies - NGC~3395 --  gave mass ratio \HI$/$H$_2\sim 21$ (see tables 11, 12 in \citealt{Boselli2014}).
If the  enriched gas is well mixed, and the enrichment due to stellar
evolution may be considered as instantaneous, then the simple closed box model
is acceptable. In this model, the oxygen abundance is connected with the
relative mass of gas  $\mu = M_{gas} / (M_{gas} + M_*) $ and with the yield
$Y_o$ of oxygen per the mass of a single generation of stars  by  the simple
relation:  $$Y_o = \frac{12(O/H)}{\ln(1/\mu)}.$$

It is convenient to introduce the effective yield, $Y_{eff}$,  as the yield
determined from the equation above for the closed box  model, which in general
may be too simple to be correct.  Accretion, as well as the outflow of gas
reduces the effective yield, so that $Y_{eff}<Y_o$ \citep{Edmunds1990,
Spitoni2010}.  Hence, a comparison of the observed estimates of $ Y_{eff}$
with the expected oxygen yield $Y_o$ allows us to reveal the peculiarities of
chemical evolution of interstellar gas in a given galaxy. Theoretically
obtained values $Y_o$ are  very uncertain because  they depend on many
factors which are difficult to take into account (see e.g. a discussion in
\citealt{Pilyugin2007}). More reliable estimates are based on the statistically
determined upper boundary of $Y_{eff}$ which is assumed to be close to $Y_o$ (
Pilyugin et al. 2007). Using this approach, Pilyugin et al. (2007) found
the most probable value $Y_0=0.035$, which agrees with \cite{Dalcanton2007} results $Y_0=
(4\pm1)\times10^{-3}$ for galaxies having the velocity of rotation $V_c> 100$
\kms. The measured values of $Y_{eff}$ for M33
(\citealt{Saburova2012}) and M51  (\citealt{Bresolin2004}) agree with these estimates for a wide range of
radial distances.

Table~\ref{t1} gives the $Y_{eff}$ estimates obtained for the oxygen abundance  \logoh
$=8.4$ for both galaxies in Arp~270 separately, and for this system as a
whole.  The latter estimate should be  the most representative one because of
the evident gas exchange between two galaxies. The data clearly demonstrate
that $Y_{eff}$ for this system is lower than expected for closed box model.
If to apply this model using $Y_o=0.0035$, then the ratio $M_{gas}/M_*$ for
Arp 270 should be about 42\%, which is much higher than the observed ratio
(see Table 1). Note that the $T_e$ method gives (although uncertain) even
lower (O/H) ratio, which, if to accept it, will  make a discrepancy with the
closed box model even deeper.

Most likely, the low $Y_{eff}$ for Arp~270 is the result of significant gas losses
experienced by the system during the previous convergence(s) of galaxies. The
lost gas should be hot and ionized, otherwise it would be visible in \HI-line
as neutral intergalactic gas. Indeed, a neutral gas is  observed in the
remnant  of tidal tails in Arp~270, but it contains only a small fraction of
the present-day \HI\ in galaxies. Is the bulk of lost gas ionized?

Note that the ionized gas mixing as the mechanism of levelling off the abundance
gradient agrees with the expected short time of gas exchange between galaxies,
although the details of this process remain not clear. Diffuse X-ray gas is
really presented in Arp~270, as the {\it Chandra} observatory shows; however, it is
observed within the optical borders of galaxies only (\citealt{Brassington2005}):
the gas outflow may be developed at later stage of interaction
(\citealt{Brassington2007}).   Another way to account for the low gradients of
metallicity is to assume that the accretion of enriched extragalactic gas on to
the discs of galaxies took place after their formation. To explain the
observed O/H ratio, the accreted gas should possess about  1/2 - 1/3 of solar
metallicity, which is typical for gas in clusters of galaxies independently on
their total stellar masses (\citealt{Renzini2014}). Although Arp~270 is quite
isolated interacting system, it is located at the periphery of NGC~3430 group,
being its possible member (the distance between Arp~270 and NGC~3430 is about
29 arcmin or 180 kpc).  However, there is no evidence of the intracluster gas in
this group, which makes the latter version doubtful.

\subsection{Ages of blue stellar islands}

Both slits passed through several blue islands of star formation, easily
visible in the images of Arp~270 both on the emission and in the continuum.
The brightest emission region crossed by Slit \No1 is ESC (see above). It is the
site of the active star formation, containing several (at least two) stellar
agglomerates within the kpc-size region. The emission spectrum of ESC does not
differ significantly from the spectra of other \HII\ regions by the line
intensity ratios.  It also does not stand out by its abundance or by
kinematics, being close to the adjacent regions of NGC~3995 by its line-of-sight 
velocity. Hence, the stellar island was formed from the gas belonging to
this galaxy, rather than appeared as a result of accretion or minor merging.

\begin{figure}
\centerline{
\includegraphics[width=0.5\textwidth]{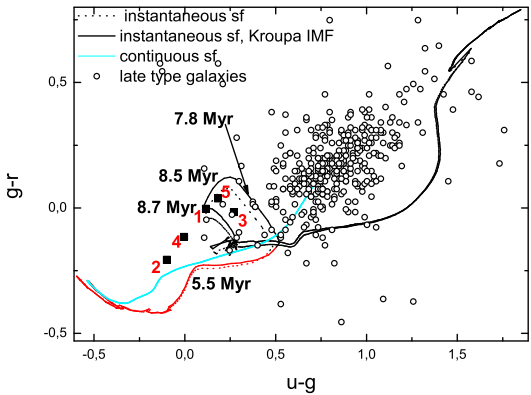}
}
\caption{The $(g-r)-(u-r)$ diagram. Squares denote the positions of bright emission
regions of Arp~270 (1 corresponds to ESC). Black lines correspond  to Starburst99 model tracks
obtained for IMF described by two exponents (see text). Blue line  corresponds
to the Starburst99 model for continuous star formation. Open circles mark the
positions of integrated colours of bright ($B<13^m$) irregular galaxies
(morphological type $T_{morph}>7$ according to HyperLeda data base). Red line marks the
region of the diagram where the influence of emission lines on the colour
indices may be significant. }
\label{colours}
\end{figure}

Fig.~\ref{colours} presents the  positions of several bright emission regions
of Arp~270 including ESC (1) on the  two-colour diagram $(g-r)-(u-r)$ (see
Fig.~\ref{arp270_image} for their location). The data for these regions were
obtained from the photometry based on the SDSS sky survey images after
correction for extinction. For comparison, the positions of integrated colours
of bright ($B<13^m$) irregular galaxies  (morphological type $T_{morph}>7$)
are given, obtained from the observed UBV colours taken from the HyperLeda
data base. A high dispersion of the total  colour indices is mainly due to
irregular character of star formation in the late-type galaxies and due to 
the estimate errors. The
colour indices of chosen \HII\ regions galaxies are bluer than those for  the
Irr galaxies because the latter contain  a mixture of old and young stellar
populations, whereas old stars are absent if we observe the beginning of star
formation.  To estimate the ages of stars, we used Starburst99 evolution
tracks for different histories of star formation: instantaneous star formation
with the IMFs described by two  exponents of 1.3 and 2.3  (solid black line)
and  1.3 and 2.0  (dash black line) for low (from 0.1 to 0.5 $M\odot$) and high (from 0.5 to 100 $M\odot$) masses, respectively,  and
for continuous star formation (blue line). Each point of the tracks corresponds to the different age of stellar population. For instantaneous and continuous star formation, we considered the following ranges of age of stellar population respectively: $1\times 10^{-2}$ Myr - 10 Gyr, $1\times 10^{-11}$ Myr - 10 Gyr. We also mark in the Fig.~\ref{colours} the age of stellar population which corresponds to the closest point of the track to the considered regions of star formation. Red line shows the region of
extremely young population ($T<6$ Myr)  where the influence of emission lines
on the colour indices which here we ignore may be significant (\citealt{Smith2008}).
As the diagram shows, star formation in all considered regions has begun
recently. The regions 2 and 4 are especially young, and their positions fit
the evolution tracks not so good (colour indices of young population strongly
depend on the heavy part of IMF and on the presence of emission lines). In
turn, the regions ESC (1), 5 and 3 consist of stars formed $8-9$ Myr ago without
the evidences of the presence of old stars.

Applying this result to ESC, we may conclude, that star formation in this
extended stellar island began recently. There may be two scenarios of
formation of this stellar island. First, the massive gaseous complex might
already exist in NGC~3395 before the convergence of galaxies, so that the
tidal force swept it out together with more rarefied gas. In this case, ESC may
be considered as gravitationally bound and long-living gaseous agglomerate
with the  star formation delayed until recent epoch, so it has a chance to
become a tidal dwarf galaxy. This way of evolution of massive gas clouds was
discussed in \cite{Zasov2014}. However, the line-of-sight velocity distribution
along ESC does not give any evidence of its kinematic decoupling which casts
doubt on this interpretation. Secondly, ESC may be formed as the result of the
compression of gas in the colliding  large-scale  gaseous flows induced by
close interaction between galaxies. It agrees with the position of ESC  in the
transition region where the line-of-sight velocities of gas  begin steeply
change from the values which may  be ascribed to NGC~3395 to those
corresponding to NGC~3396.

\section{General conclusions}

\begin{itemize} 

\item We obtain the line-of-sight velocities of gas and stars
and chemical abundances distributions along the two slits crossing the
peripheral regions of the interacting galaxies. Irregular velocity distribution
and the high velocity dispersion of emitting gas which stems from the profiles
of emission lines evidences the  non-circular gas flows at small scales up
to $\approx1$ kpc. We found that the stellar component of NGC~3395 differs by its velocity from the emission gas component at about 50 \kms\ in the extended region cut by the slit in the periphery of NGC~3395, which evidences a spatial separation of stars and  gas in the tidally disturbed galaxy.

\item The ratio of emission lines along the slits evidences that they are
produced by non-resolved \HII\ regions. Their velocity dispersion is high ($\sim50$
\kms) in comparison with normal star-forming galaxies.

\item The abundance estimates of oxygen obtained by different methods reveal
the absence of significant gradients of metallicity. The oxygen abundance is
found as \logoh $\approx 8.4$, that is the gas metallicity of about a half of
solar value. Either there was an effective mixing of gas inside of galaxies
and between them, or (less probable), we observe the result of  accretion of
initially enriched gas on to the galaxies.

\item Assuming a constant (O/H) along the system, we demonstrate that the
system is badly described by the closed box model: the effective yield
of oxygen  $Y_{eff}$ is significantly lower than expected one for closed box
model, although 3D spectroscopy is needed to strengthen this conclusion.
 Low  $Y_{eff}$ may be explained by gas losses experienced by the system during
the previous convergence(s) of galaxies.

\item Colour indices ($ugr$) of several discrete sites of star formation beyond
the inner regions of galaxies were estimated from the SDSS data. They
confirmed the young age ($T<10^7$ yr) of stellar population. The extended kpc
size island of star formation (ESC) observed between the galaxies does not stand out
by its kinematics or abundances, so it hardly may be considered as the tidal
dwarf candidate. Its location at the beginning of transition zone between
galaxies allows us to propose that its formation is the result of compression of
colliding gas flows of galaxies in contact.
	
\end{itemize}

\section*{Acknowledgements} The observations at the 6-meter BTA telescope were carried out
with the  financial support of the Ministry of Education and Science of the Russian Federation (agreement No. 14.619.21.0004, project ID RFMEFI61914X0004). We are grateful to the anonymous referee for the valuable comments  and advices. In this study, we used the SDSS DR9 data. Funding
for the SDSS and SDSS-II has been provided by the Alfred P. Sloan Foundation,
the Participating Institutions, the National Science Foundation, the U.S.
Department of Energy, the National Aeronautics and Space Administration, the
Japanese Monbukagakusho, the Max Planck Society, and the Higher Education
Funding Council for England. The SDSS website is http://www.sdss.org/. This research has made use of the NASA/IPAC Extragalactic Data base (NED) which is operated by the Jet Propulsion Laboratory, California Institute of Technology, under contract with the National Aeronautics and Space Administration.  
We acknowledge the use of the HyperLeda data base. This work was supported by 
Russian Foundation for Basic Research (projects Nos. 12-02-00685 and 14-22-03006-ofi-m). 
IYK is also grateful to Dmitry Zimin's non-profit Dynasty Foundation.

\bibliographystyle{mn2e}
\bibliography{arp270}

\begin{thebibliography}{47}
\expandafter\ifx\csname natexlab\endcsname\relax\def\natexlab#1{#1}\fi

\bibitem[{{Afanasiev} \& {Moiseev}(2011)}]{scorpio2}
{Afanasiev} V.~L., {Moiseev} A.~V., 2011, Baltic Astronomy, 20, 363

\bibitem[{{Asplund} {et~al}\mbox{.}(2009){Asplund}, {Grevesse}, {Sauval}, \&
  {Scott}}]{Asplund2009}
{Asplund} M., {Grevesse} N., {Sauval} A.~J., {Scott} P., 2009, \araa, 47, 481

\bibitem[{{Bell} \& {de Jong}(2001)}]{Bell2001}
{Bell} E.~F., {de Jong} R.~S., 2001, \apj, 550, 212

\bibitem[{{Bell} {et~al}\mbox{.}(2003){Bell}, {McIntosh}, {Katz}, \&
  {Weinberg}}]{Bell2003}
{Bell} E.~F., {McIntosh} D.~H., {Katz} N., {Weinberg} M.~D., 2003, \apjs, 149,
  289

\bibitem[{{Boselli}, {Cortese} \& {Boquien}(2014){Boselli}, {Cortese}, \&
  {Boquien}}]{Boselli2014}
{Boselli} A., {Cortese} L., {Boquien} M., 2014, \aap, 564, A65

\bibitem[{{Brassington}, {Ponman} \& {Read}(2007){Brassington}, {Ponman}, \&
  {Read}}]{Brassington2007}
{Brassington} N.~J., {Ponman} T.~J., {Read} A.~M., 2007, \mnras, 377, 1439

\bibitem[{{Brassington}, {Read} \& {Ponman}(2005){Brassington}, {Read}, \&
  {Ponman}}]{Brassington2005}
{Brassington} N.~J., {Read} A.~M., {Ponman} T.~J., 2005, \mnras, 360, 801

\bibitem[{{Bresolin}, {Garnett} \& {Kennicutt}(2004){Bresolin}, {Garnett}, \&
  {Kennicutt}}]{Bresolin2004}
{Bresolin} F., {Garnett} D.~R., {Kennicutt}, Jr. R.~C., 2004, \apj, 615, 228

\bibitem[{{Chilingarian} {et~al}\mbox{.}(2007{\natexlab{a}}){Chilingarian},
  {Prugniel}, {Sil'Chenko}, \& {Koleva}}]{nbursts_a}
{Chilingarian} I., {Prugniel} P., {Sil'Chenko} O., {Koleva} M.,
  2007{\natexlab{a}}, in IAU Symposium, Vol. 241, IAU Symposium, {Vazdekis} A.,
  {Peletier} R., eds., pp. 175--176

\bibitem[{{Chilingarian} {et~al}\mbox{.}(2007{\natexlab{b}}){Chilingarian},
  {Prugniel}, {Sil'Chenko}, \& {Afanasiev}}]{nbursts_b}
{Chilingarian} I.~V., {Prugniel} P., {Sil'Chenko} O.~K., {Afanasiev} V.~L.,
  2007{\natexlab{b}}, \mnras, 376, 1033

\bibitem[{{Clemens} {et~al}\mbox{.}(1999){Clemens}, {Baxter}, {Alexander}, \&
  {Green}}]{Clemens}
{Clemens} M.~S., {Baxter} K.~M., {Alexander} P., {Green} D.~A., 1999, \mnras,
  308, 364

\bibitem[{{Corbelli}(2003)}]{Corbelli2003}
{Corbelli} E., 2003, \mnras, 342, 199

\bibitem[{{Dalcanton}(2007)}]{Dalcanton2007}
{Dalcanton} J.~J., 2007, \apj, 658, 941

\bibitem[{{D'Odorico}(1970)}]{DOdorico}
{D'Odorico} S., 1970, \apj, 160, 3

\bibitem[{{Edmunds}(1990)}]{Edmunds1990}
{Edmunds} M.~G., 1990, \mnras, 246, 678

\bibitem[{{Epinat}, {Amram} \& {Marcelin}(2008){Epinat}, {Amram}, \&
  {Marcelin}}]{Ghasp}
{Epinat} B., {Amram} P., {Marcelin} M., 2008, \mnras, 390, 466

\bibitem[{{Garrido} {et~al}\mbox{.}(2002){Garrido}, {Marcelin}, {Amram}, \&
  {Boulesteix}}]{Garrido}
{Garrido} O., {Marcelin} M., {Amram} P., {Boulesteix} J., 2002, \aap, 387, 821

\bibitem[{{Hancock} {et~al}\mbox{.}(2003){Hancock}, {Weistrop}, {Eggers}, \&
  {Nelson}}]{Hancock}
{Hancock} M., {Weistrop} D., {Eggers} D., {Nelson} C.~H., 2003, \aj, 125, 1696

\bibitem[{{Izotov} {et~al}\mbox{.}(2006){Izotov}, {Stasi\'nska}, {Meynet},
  {Guseva}, \& {Thuan}}]{Izotov2006}
{Izotov} Y., {Stasi\'nska} G., {Meynet} G., {Guseva} N., {Thuan} T., 2006,
  \aap, 448, 955

\bibitem[{{Katkov} \& {Chilingarian}(2011)}]{skysubtraction}
{Katkov} I.~Y., {Chilingarian} I.~V., 2011, in Astronomical Society of the
  Pacific Conference Series, Vol. 442, Astronomical Data Analysis Software and
  Systems XX, {Evans} I.~N., {Accomazzi} A., {Mink} D.~J., {Rots} A.~H., eds.,
  p. 143

\bibitem[{{Kewley} {et~al}\mbox{.}(2006){Kewley}, {Groves}, {Kauffmann}, \&
  {Heckman}}]{Kewley2006}
{Kewley} L.~J., {Groves} B., {Kauffmann} G., {Heckman} T., 2006, \mnras, 372,
  961

\bibitem[{{Kroupa}(2002)}]{Kroupa2002}
{Kroupa} P., 2002, Science, 295, 82

\bibitem[{{Leitherer} {et~al}\mbox{.}(1999){Leitherer}, {Schaerer}, {Goldader},
  {Delgado}, {Robert}, {Kune}, {de Mello}, {Devost}, \&
  {Heckman}}]{Leitherer1999_sb99}
{Leitherer} C. {et~al.}, 1999, \apjs, 123, 3

\bibitem[{{Libeskind} {et~al}\mbox{.}(2011){Libeskind}, {Knebe}, {Hoffman},
  {Gottl{\"o}ber}, \& {Yepes}}]{Libeskind2011}
{Libeskind} N.~I., {Knebe} A., {Hoffman} Y., {Gottl{\"o}ber} S., {Yepes} G.,
  2011, \mnras, 418, 336

\bibitem[{{Magrini} {et~al}\mbox{.}(2010){Magrini}, {Stanghellini}, {Corbelli},
  {Galli}, \& {Villaver}}]{Magrini2010}
{Magrini} L., {Stanghellini} L., {Corbelli} E., {Galli} D., {Villaver} E.,
  2010, \aap, 512, A63

\bibitem[{{Makarov} {et~al}\mbox{.}(2014){Makarov}, {Prugniel}, {Terekhova},
  {Courtois}, \& {Vauglin}}]{Makarov2014}
{Makarov} D., {Prugniel} P., {Terekhova} N., {Courtois} H., {Vauglin} I., 2014,
  \aap, 570, A13

\bibitem[{{Martins} {et~al}\mbox{.}(2005){Martins}, {Gonz{\'a}lez Delgado},
  {Leitherer}, {Cervi{\~n}o}, \& {Hauschildt}}]{Martins2005}
{Martins} L.~P., {Gonz{\'a}lez Delgado} R.~M., {Leitherer} C., {Cervi{\~n}o}
  M., {Hauschildt} P., 2005, \mnras, 358, 49

\bibitem[{{Miralles-Caballero} {et~al}\mbox{.}(2014){Miralles-Caballero},
  {D{\'i}az}, {Rosales-Ortega}, {P{\'e}rez-Montero}, \&
  {S{\'a}nchez}}]{Miralles2014}
{Miralles-Caballero} D., {D{\'i}az} A.~I., {Rosales-Ortega} F.~F.,
  {P{\'e}rez-Montero} E., {S{\'a}nchez} S.~F., 2014, \mnras, 440, 2265

\bibitem[{{Moiseev}, {Tikhonov} \& {Klypin}(2014){Moiseev}, {Tikhonov}, \&
  {Klypin}}]{Moiseev2014}
{Moiseev} A.~V., {Tikhonov} A.~V., {Klypin} A., 2014, ArXiv e-prints:1405.5731

\bibitem[{{Pilyugin} {et~al}\mbox{.}(2009){Pilyugin}, {Mattsson}, {V\'ilchez},
  \& {Cedr\'es}}]{Pilyugin2009}
{Pilyugin} L., {Mattsson} L., {V\'ilchez} J., {Cedr\'es} B., 2009, \mnras, 398,
  485

\bibitem[{{Pilyugin}, {V\'ilchez} \& {Thuan}(2010){Pilyugin}, {V\'ilchez}, \&
  {Thuan}}]{Pilyugin2010}
{Pilyugin} L., {V\'ilchez} J., {Thuan} T., 2010, \apj, 720, 1738

\bibitem[{{Pilyugin} \& {Mattsson}(2011)}]{Pilyugin2011}
{Pilyugin} L.~S., {Mattsson} L., 2011, \mnras, 412, 1145

\bibitem[{{Pilyugin}, {Thuan} \& {V{\'i}lchez}(2007){Pilyugin}, {Thuan}, \&
  {V{\'i}lchez}}]{Pilyugin2007}
{Pilyugin} L.~S., {Thuan} T.~X., {V{\'i}lchez} J.~M., 2007, \mnras, 376, 353

\bibitem[{{Renzini} \& {Andreon}(2014)}]{Renzini2014}
{Renzini} A., {Andreon} S., 2014, \mnras, 444, 3581

\bibitem[{{Rupke}, {Kewley} \& {Barnes}(2010){Rupke}, {Kewley}, \&
  {Barnes}}]{Rupke2010}
{Rupke} D.~S.~N., {Kewley} L.~J., {Barnes} J.~E., 2010, \apjl, 710, L156

\bibitem[{{Saburova} \& {Zasov}(2012)}]{Saburova2012}
{Saburova} A.~S., {Zasov} A.~V., 2012, Astronomy Letters, 38, 139

\bibitem[{{S{\'a}nchez} {et~al}\mbox{.}(2014){S{\'a}nchez}, {Rosales-Ortega},
  {Iglesias-P{\'a}ramo}, {Moll{\'a}}, {Barrera-Ballesteros}, {Marino},
  {P{\'e}rez}, {S{\'a}nchez-Blazquez}, {Gonz{\'a}lez Delgado}, {Cid Fernandes},
  {de Lorenzo-C{\'a}ceres}, {Mendez-Abreu}, {Galbany}, {Falcon-Barroso},
  {Miralles-Caballero}, {Husemann}, {Garc{\'{\i}}a-Benito}, {Mast}, {Walcher},
  {Gil de Paz}, {Garc{\'{\i}}a-Lorenzo}, {Jungwiert}, {V{\'{\i}}lchez},
  {J{\'{\i}}lkov{\'a}}, {Lyubenova}, {Cortijo-Ferrero}, {D{\'{\i}}az},
  {Wisotzki}, {M{\'a}rquez}, {Bland-Hawthorn}, {Ellis}, {van de Ven}, {Jahnke},
  {Papaderos}, {Gomes}, {Mendoza}, \& {L{\'o}pez-S{\'a}nchez}}]{Sanchez2014}
{S{\'a}nchez} S.~F. {et~al.}, 2014, \aap, 563, A49

\bibitem[{{Smith} {et~al}\mbox{.}(2010){Smith}, {Giroux}, {Struck}, \&
  {Hancock}}]{Smith}
{Smith} B.~J., {Giroux} M.~L., {Struck} C., {Hancock} M., 2010, \aj, 139, 1212

\bibitem[{{Smith} {et~al}\mbox{.}(2008){Smith}, {Struck}, {Hancock}, {Giroux},
  {Appleton}, {Charmandaris}, {Reach}, {Hurlock}, \& {Hwang}}]{Smith2008}
{Smith} B.~J. {et~al.}, 2008, \aj, 135, 2406

\bibitem[{{Spitoni} {et~al}\mbox{.}(2010){Spitoni}, {Calura}, {Matteucci}, \&
  {Recchi}}]{Spitoni2010}
{Spitoni} E., {Calura} F., {Matteucci} F., {Recchi} S., 2010, \aap, 514, A73

\bibitem[{{van der Marel} \& {Franx}(1993)}]{gausshermite}
{van der Marel} R.~P., {Franx} M., 1993, \apj, 407, 525

\bibitem[{{van Dokkum}(2001)}]{lacosmic}
{van Dokkum} P.~G., 2001, \pasp, 113, 1420

\bibitem[{{V{\'a}zquez} \& {Leitherer}(2005)}]{Vazquez2005_sb99}
{V{\'a}zquez} G.~A., {Leitherer} C., 2005, \apj, 621, 695

\bibitem[{{Werk} {et~al}\mbox{.}(2011){Werk}, {Putman}, {Meurer}, \&
  {Santiago-Figueroa}}]{Werk2011}
{Werk} J.~K., {Putman} M.~E., {Meurer} G.~R., {Santiago-Figueroa} N., 2011,
  \apj, 735, 71

\bibitem[{{Werk} {et~al}\mbox{.}(2010){Werk}, {Putman}, {Meurer}, {Thilker},
  {Allen}, {Bland-Hawthorn}, {Kravtsov}, \& {Freeman}}]{Werk2010}
{Werk} J.~K., {Putman} M.~E., {Meurer} G.~R., {Thilker} D.~A., {Allen} R.~J.,
  {Bland-Hawthorn} J., {Kravtsov} A., {Freeman} K., 2010, \apj, 715, 656

\bibitem[{{Zaragoza-Cardiel} {et~al}\mbox{.}(2013){Zaragoza-Cardiel},
  {Font-Serra}, {Beckman}, {Blasco-Herrera}, {Garc{\'i}a-Lorenzo}, {Camps},
  {Gonzalez-Martin}, {Ramos Almeida}, {Loiseau}, \&
  {Guti{\'e}rrez}}]{Zaragoza-Cardiel}
{Zaragoza-Cardiel} J. {et~al.}, 2013, \mnras, 432, 998

\bibitem[{{Zasov} \& {Kasparova}(2014)}]{Zasov2014}
{Zasov} A., {Kasparova} A., 2014, \apss, 353, 595

\end{thebibliography}

\label{lastpage}
\end{document}